# A Strongly Miniaturized and Inherently Matched Folded Dipole Antenna for Narrowband Applications

Sanghamitro Das, *Member, IEEE*, David J. Sawyer, *Member, IEEE*, Nectaria Diamanti,
A. Peter Annan, and Ashwin K. Iyer, *Senior Member, IEEE*

*Abstract*—A novel miniaturized printed folded dipole (FD) antenna has been designed for narrowband sensing applications. It is shown that the antenna may be systematically engineered to achieve matching to a practical source impedance at frequencies well below its half-wavelength resonance using series-*LC* loading and asymmetrical arm widths. This approach is used to design an inherently well-matched antenna that does not require an external matching network and demonstrates approximately 75% miniaturization, excellent co- to cross-polarization separation, over 24 dB higher realized gain, and strongly improved radiation efficiency relative to an unloaded FD of the same electrically small size. Experimental validation of a fabricated prototype demonstrates excellent agreement with simulations.

*Index Terms*—Folded dipole (FD) antenna, miniaturization, reactive loading.

## I. INTRODUCTION

THE process of miniaturizing an antenna, which is often the largest component in a wireless system, generally degrades its efficiency and/or bandwidth significantly. Nevertheless, electrically small antennas are extensively used in low-power sensing/monitoring applications, such as RFID tags, biomedical implant antennas, or Internet of Things (IoT) devices [1]–[6], where radiation efficiency and bandwidth are often sacrificed for the sake of compactness. In such antennas, the focus turns toward increasing the power accepted by the antenna, which invokes impedance-matching techniques. An increase in accepted power implies enhanced radiated power for a given radiation efficiency and, hence, a longer detection range for an electrically small sensor antenna.

Manuscript received July 15, 2018; revised October 4, 2019; accepted December 15, 2019. Date of publication January 7, 2020; date of current version May 5, 2020. This work was supported in part by the Natural Sciences and Engineering Research Council (NSERC) of Canada through a Collaborative Research and Development (CRD) Grant with Sensors & Software Inc., Mississauga, ON, Canada, and in part by a Graduate Fellowship from Alberta Innovates. *(Corresponding author: Ashwin Iyer.)*

Sanghamitro Das and Ashwin K. Iyer are with the Department of Electrical and Computer Engineering, University of Alberta, Edmonton, AB T6G 1H9, Canada (e-mail: iyer@ece.ualberta.ca).

David J. Sawyer is with DVTEST Inc., Pickering, ON L1W 3W9, Canada.

Nectaria Diamanti is with the Department of Geophysics, Aristotle University of Thessaloniki, 541 24 Thessaloniki, Greece.

A. Peter Annan is with Sensors & Software Inc., Mississauga, ON L4W 2X8, Canada.

The challenge, of course, is that reduction in the antenna size typically results in a small input resistance and a large input reactance, making it very difficult to match to a practical source impedance. The use of external matching networks to mitigate this problem often comes at the expense of increasing the size and complexity of the structure.

Some of the most widely used antennas for sensing applications are dipole antennas and their folded-dipole counterparts, particularly because of their simplicity, low cost, compact size, well-defined radiation properties, omnidirectional coverage, and simple analysis and design. It is well known that conventional dipole antennas can be miniaturized using various reactive loading techniques [7]–[12]. Similar techniques have been employed for folded dipole (FD) antennas, including loading with a series inductance [13] or a shunt capacitance (also known as "top-loading") [14]–[15]; or using both simultaneously [16]. Furthermore, external *LC* resonators were used to miniaturize folded antennas, where the operating frequency was determined by the resonance frequency of the loading inductors and capacitors, and not by the length of the antenna [17].

More recently, electrically small monopole and dipole antennas have been realized by applying metamaterial loading [18]–[20] or parasitic metamaterial-inspired elements in close proximity [21]–[22]. Metamaterial technology has also been exploited for the miniaturization of FD antennas. Electrically small but efficient folded antennas were realized by forcing the antenna arms to radiate in phase using either conventional or metamaterial phase-shifting lines [23]–[25] or by zero-degree negative-refractive-index transmission-line (NRI-TL) metamaterial unit cells [26].

To appreciate the constraints on dipole miniaturization, we may consider the input-impedance responses $Z_{in}(f)$ of both conventional and FDs. These are shown in Fig. 1(a) and (b) for slender dipoles in vacuum. The aforementioned miniaturization methods focus only on $f_R$, which is the resonance frequency at which the linear dimension of the antenna is $l \approx 0.5\lambda$. However, lowering $f_R$ using any of these miniaturization techniques decreases the input resistance $\Re\{Z_{in}(f_R)\}$ dramatically, making the antenna unsuitable for matching to a practical source impedance. The traditional way to address this problem is to use an external matching network. The use of FDs alleviates this constraint somewhat, since the FD antenna provides a higher $\Re\{Z_{in}(f_R)\}$ than its conventional counterpart, which may be increased further by employing multiple folded arms [16], [26]–[29] or by creating asymmetries in the arm widths [13]–[15], [30]. Unfortunately, such techniques generally increase complexity and the footprint of these antennas by requiring very high arm-width ratios, which may be counterproductive to the goal of miniaturization. An alternative approach to embedded matching involves the inclusion of non-Foster matching elements inside the antenna structure, which has been studied for conventional dipole/monopole and patch antennas [31]–[32]. However, these techniques typically employ multiple reactive elements at each loading location. Additionally, an external dc bias will always be required for implementing such non-Foster networks. As a result, these methods have not seen wide adoption beyond lower UHF frequencies, i.e., a few hundred megahertz, mainly due to the scarcity of available physical space in the extreme electrically small regimes, where the implementation of such embedded non-Foster loading elements becomes difficult. Furthermore, non-Foster elements typically suffer from issues like stability, nonlinearity, and loss, which also prevent them from being used in many of the very applications for which they would be most useful. As such, a purely passive loading scheme that keeps the number of loading elements to a minimum would be much more desirable for designing narrowband miniaturized antennas operating near and above gigahertz frequencies.

A distinguishing feature of the FD antenna is that, unlike its conventional counterpart, the resonance frequency $f_R$ is surrounded by two antiresonances $f_{AR,1}$ and $f_{AR,2}$, as can be seen through comparison of Fig. 1(a) and (b). Although $f_{AR,1}$ is always lower than $f_R$, the very high value of $\Re\{Z_{in}(f_{AR,1})\}$ once again makes it unamenable for impedance matching to practical sources. Moreover, $f_{AR,1}$ exhibits lower radiation efficiency in comparison with $f_R$, since the FD currents at $f_{AR,1}$ do not radiate in phase as they do at $f_R$. As a result, $f_{AR,1}$ is not considered a useful operating frequency for an FD antenna. However, the use of $f_{AR,1}$ can be justified, provided the issue of very high $\Re\{Z_{in}(f_{AR,1})\}$ can somehow be resolved, so that good matching, and therefore higher radiated power, can be ensured.

The main contribution of this article is to show that, by employing a simple (passive) reactive loading scheme and then judiciously choosing the values of the required loading reactances, the input-impedance response of an FD antenna can be practically engineered in order to achieve good impedance matching at a desired miniaturization level with respect to an arbitrary source impedance, without the need for external matching networks or other techniques that increase the overall footprint of the antenna and/or its complexity. Specifically, it will be shown that a series combination of inductive and capacitive lumped loading introduces a new resonance frequency below $f_{AR,1}$, and that this new resonance together with $f_{AR,1}$ may be used to control the $Z_{in}$ of the antenna in order to achieve excellent return loss and better radiation efficiency compared to an unloaded FD antenna of the same size at the same frequency. An asymmetric reactive loading scheme is used to realize a greater current imbalance between the arms of the FD in order to improve the radiation efficiency. Finally, the use of unequal arm widths has been exploited as an additional degree of freedom in conjunction with the reactive loading in order to fine-tune the impedance matching and the radiation performance. It will be shown that a reasonably small arm-width ratio will be required for this case, thus a small antenna footprint can be ensured.

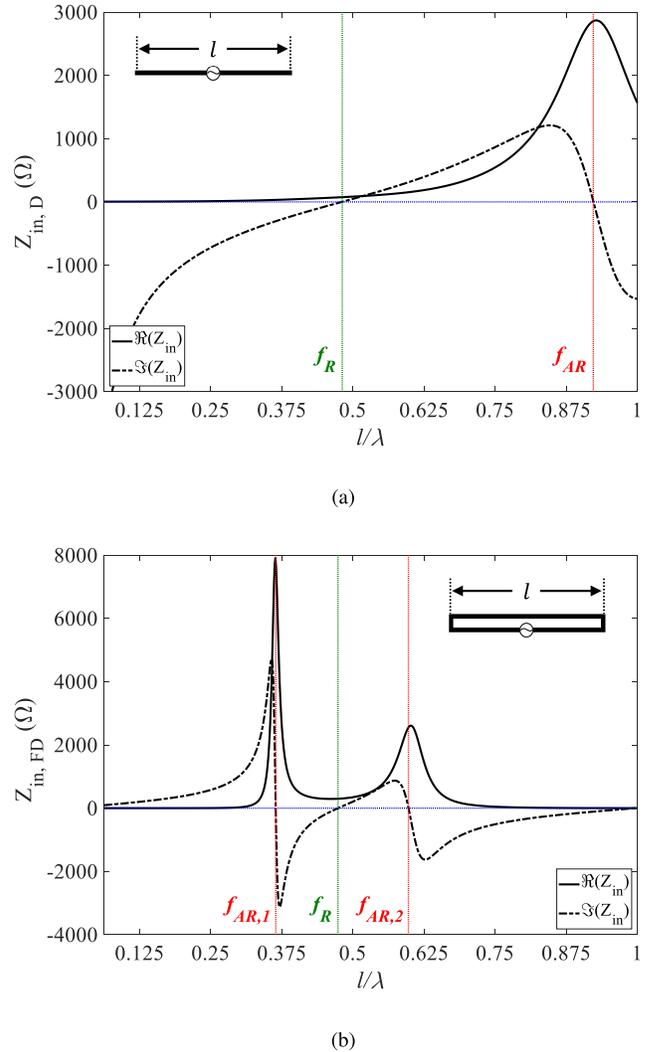

Fig. 1. Input-impedance profiles of (a) conventional dipole (D) antenna and (b) FD antenna.

This article is organized as follows. Section II presents a study on the input-impedance trends of a reactively loaded FD antenna, obtained by performing a series of full-wave parametric simulations. The major inferences drawn from the study are then exploited to design a 75% miniaturized antenna that demonstrates good return loss at the target operating frequency of 1 GHz. Several practical issues related to the components used in designing the miniaturized antenna are also discussed. In Section III, the fabrication of a prototype based on the final simulated design is described, followed by the measured results that are found to be in excellent agreement with the corresponding simulated results.

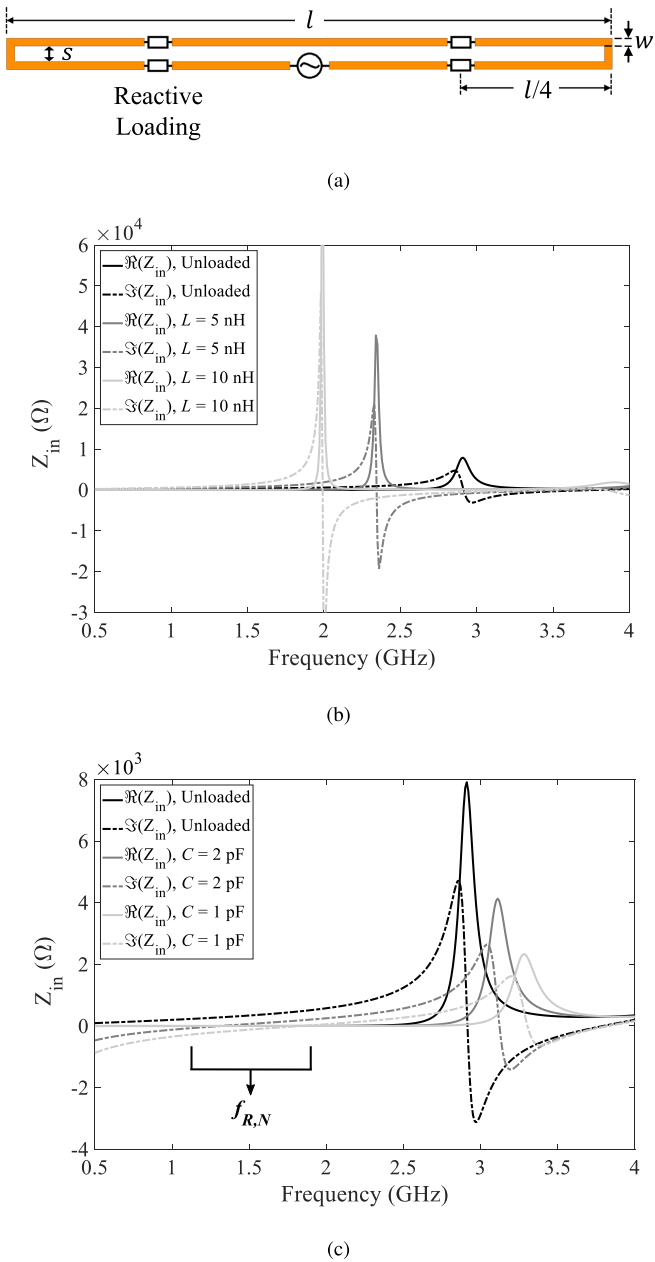

Fig. 2. (a) Schematic of the reactively loaded FD antenna (not to scale); input-impedance trends for (b) $L$-loading and (c) $C$-loading.

## II. DESIGN AND SIMULATION

### A. Effect of Reactive Loading

In order to engineer the impedance-matching response of the FD antenna, we need to first study the effects of different series inductive ($L$) and capacitive ($C$) loading values on $Z_{in}$. As a starting point, an infinitesimally thin ($w, s \ll \lambda$) planar FD antenna in vacuum is selected. A schematic of the antenna shown in Fig. 2(a) depicts the geometry, including the arm length $l$ and the loading locations. The length of the antenna was chosen to be $l = 38$ mm ($\sim 0.5\lambda$ at 3.95 GHz). The loading reactances were placed on both driven and folded arms at a distance $l/4$ from the ends, and the model was simulated using Ansys HFSS. The simulated $Z_{in}$ profiles for inductive ($L$-) and capacitive ($C$-) loading are shown in Fig. 2(b) and (c), respectively. It can be seen from Fig. 2(b) that series-$L$ loading shifts all resonances toward lower frequencies while increasing the $\Re\{Z_{in}(f_{AR,1})\}$, thus making these resonances even less suitable for matching. As shown in Fig. 2(c), the trend is expectedly the opposite for series-$C$ loading. However, it can also be noticed that the series-$C$ loading introduces a new resonance $f_{R,N}$ before $f_{AR,1}$, which is essentially the series-$LC$ resonance between the loading capacitance and the antenna inductance below $f_{AR,1}$, and is clearly at a much lower frequency than the usual half-wavelength resonance point ($f_R$). It is usually not possible to achieve good impedance matching at $f_{R,N}$, since $\Re\{Z_{in}(f_{R,N})\}$ is very small, which is opposite to the problem observed for $\Re\{Z_{in}(f_{AR,1})\}$. We now propose a novel technique whereby the value of $\Re\{Z_{in}(f_{R,N})\}$ may be increased simply by causing $f_{R,N}$ and $f_{AR,1}$ to approach each other, such that $\Re\{Z_{in}(f_{R,N})\}$ can benefit from the quickly rising slope just before $f_{AR,1}$. Indeed, as will be shown in Section II-B, $\Re\{Z_{in}(f_{R,N})\}$ and $\Re\{Z_{in}(f_{AR,1})\}$ can be engineered to the levels where good matching can be enabled with respect to a given practical source impedance by employing both inductive and capacitive loading in series, and adjusting their values intelligently. Although beyond the scope of this article, a theoretical study of the proposed reactive-loading scheme confirms that the separation between the two resonance frequencies decreases for large values of $L$ and small values of $C$. These observations are applied in Section II-B to design a practical miniaturized FD antenna.

### B. Design Strategy

A 40 mil-thick Megtron 4 R-5725 Laminate ($\varepsilon_r = 4.14$, $\tan\delta = 0.005$) dielectric substrate has been considered for this work. The antenna length $l$ was kept unchanged at 38 mm, but the arm widths and the separation between them were made larger in order to facilitate fabrication. The driven and folded arm widths were made unequal for generality and assigned two different width parameters: $w_{dr}$ and $w_{fd}$, respectively. The antenna is fed by a balanced coplanar stripline (CPS) that comprises a linearly tapered transition from a 200 $\Omega$ CPS line to the 0.5 mm-wide feed gap on the driven arm of the FD, as shown in Fig. 3(a). The traces were assigned a thickness of 35 $\mu$m and a surface roughness of 0.2 $\mu$m in order to model practical losses in the copper conductor. In order to use the available space more efficiently, the $L$ and $C$ loading elements were placed on the folded and driven arms, respectively, at the same $l/4$ distance from the ends of the antenna. This particular arrangement exploits the benefits of current imbalance between the two arms, which is expected to result in a better radiation efficiency. The loading inductors were modeled using lossless lumped models and the loading capacitors were realized in interdigitated form with gap width $C_g$ and finger length $C_l$. These loading reactances, along with the individual arm widths, were varied parametrically targeting a good impedance match at 1 GHz ($\sim 75\%$ miniaturized) with respect to the 200 $\Omega$ port impedance. It was found that for separation $s = 0.25$ mm, $L = 42.97$ nH, capacitor dimensions $C_g = 0.1$ mm, $C_l = 0.14$ mm, $C_w = 0.15$ mm, and $C_e = 0.075$ mm, driven arm width $w_{dr} = 1$ mm, and folded arm width $w_{fd} = 2$ mm, very good return loss could be obtained exactly at 1 GHz. The equivalent capacitance of the interdigitated structure was estimated using simulation to be approximately 52.5 fF. The impedance profile and $S_{11}$ of the design are shown in Fig. 3(b) and (c), respectively. It can be noticed from Fig. 3(b) that, due to the application of high loading inductance and small loading capacitance values, the resonances $f_{R,N}$ and $f_{AR,1}$ are now situated very close to each other with only 8 MHz separation between them. This results in $\Re\{Z_{in}(f_{R,N})\}$ very close to the desired 200 Ω and, thus, validates the proposed approach. The miniaturized FD antenna retains an omnidirectional radiation pattern and presents a gain of $-0.9$ dBi with 72.3% radiation efficiency, as compared to $-12.2$ dBi gain and 5.8% radiation efficiency for an unloaded FD of the same size at the same frequency. The improvement of the gain and the radiation efficiency may be inferred from the current distributions on the two antennas, shown in Fig. 4. Both current distributions are plotted on the same scale at a frequency of 1 GHz, and the arrows suggest the observed current directions. It can be easily seen that for the unloaded case, the current magnitudes on the two arms are contra-directed on either side of the gap and of almost the same magnitude, resulting in an effective cancellation of radiated fields. However, for the reactively loaded case, although the currents on the two arms remain contra-directed, the magnitude is far higher on the folded arm, which essentially produces a net unbalanced current that ultimately results in radiation, and hence, a better radiation efficiency could be realized. Finally, the simulated 10 dB return-loss bandwidth of the miniaturized antenna is expectedly narrow, measuring 7 MHz.

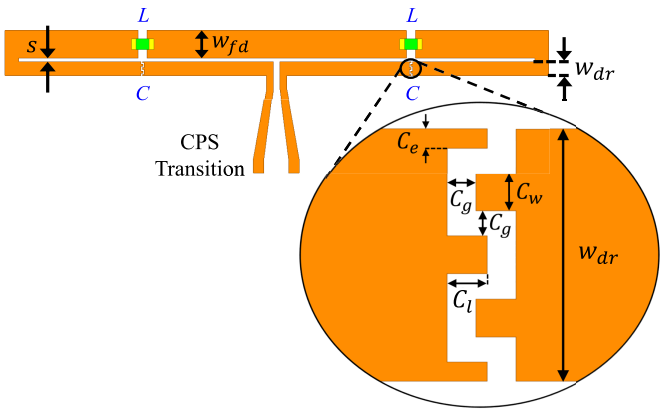

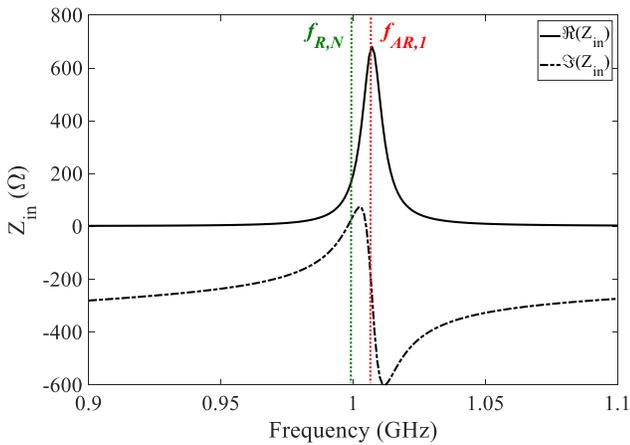

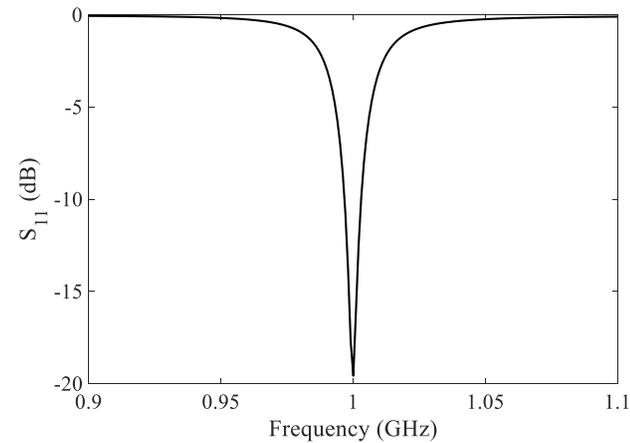

Fig. 3. (a) Reactively loaded FD antenna following the design procedure described in Section II-B. Corresponding (b) $Z_{in}$ and (c) $S_{11}$.

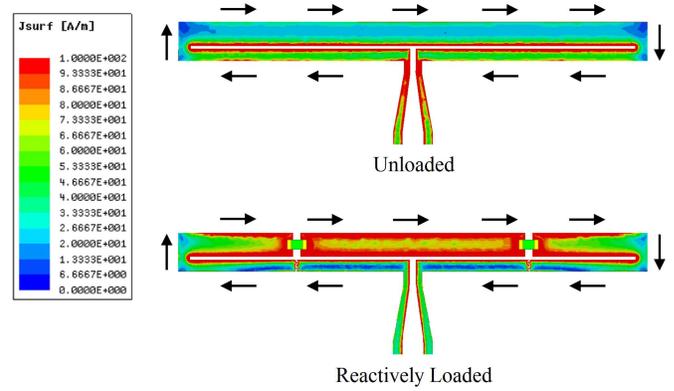

Fig. 4. Comparison of current distributions for the unloaded and reactively loaded FD antennas at 1 GHz.

Thus, although it is possible to achieve good impedance matching and radiation performance for an FD antenna at generally any miniaturization level following this method, more extreme miniaturization implies increasingly narrow bandwidths, lower radiation efficiencies, and typically more extreme values for the loading reactances. Generally, the design procedure for achieving good impedance matching at a high degree of miniaturization can be summarized as follows.

1) Choose a target frequency of operation or, conversely, the desired degree of miniaturization for a given physical length.
2) Introduce high series $L$-loading on the folded arm in order to bring $f_{AR,1}$ below the target frequency.

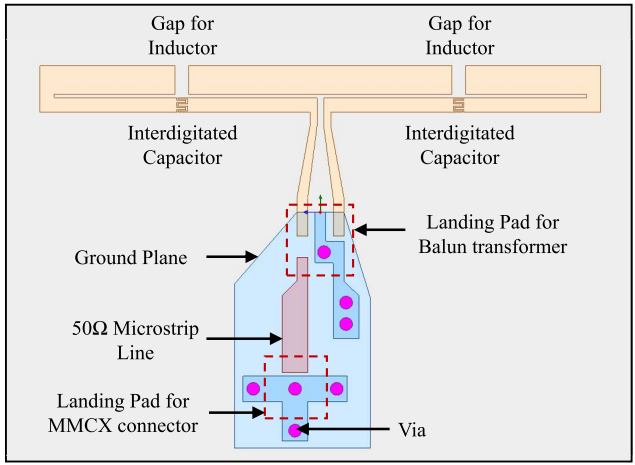

Fig. 5. Layout of miniaturized antenna with feed structure.

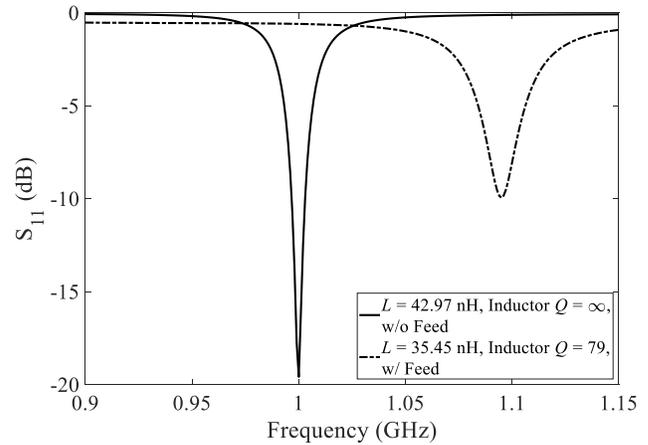

Fig. 6. Comparison of the simulated $S_{11}$ data for the antenna with $L = 42.97$ nH, $Q = \infty$, and without feed structure versus with $L = 35.45$ nH, $Q = 79$, and with feed structure.

3) Introduce series $C$-loading on the driven arm to bring $f_{AR,1}$ around the target frequency and $f_{R,N}$ very close to $f_{AR,1}$.
4) Using parametric simulations, fine-tune the loading values and the arm widths simultaneously to achieve good return loss at the target frequency.

### C. Effect of Feed Structure

Since the FD antenna requires a balanced input, for measurement using a standard 50 Ω reference impedance it is necessary to design a balun structure that can transform a 50 Ω unbalanced coaxial line to a 200 Ω balanced CPS. A surface-mount balun transformer (Mini Circuits TC4–25+) has been selected for this purpose. It is a center-tap transformer with a 1:4 turns ratio and can be operated over a bandwidth of 500 MHz to 2.5 GHz. A microstrip-based feed structure has been designed to house the balun transformer and a microminiature coaxial (MMCX) surface mount connector, which can be fed via a 50 Ω MMCX cable. The total length of the feed structure is 23.5 mm (0.08$\lambda$ at 1 GHz), which is much smaller than conventional printed microstrip-to-CPS balun structures that require approximately quarter-wavelength (or even longer) transition lengths [33]–[37]. The feed structure along with the antenna is shown in Fig. 5.

In order to determine the frequency response of the feed layout, it was simulated in HFSS with 50 Ω and 200 Ω port impedances at the input and output sides, respectively. The balun transformer could not be included in the simulation due to the unavailability of an accurate full-wave model. However, the scattering parameters of the feed layout obtained using HFSS could be co-simulated with the published scattering-parameter data of the transformer [38] using the Keysight Advanced Design System (ADS) microwave circuit simulator. It was found that the feed structure has an excellent broadband (return loss better than 18 dB) and low loss (insertion loss better than 0.2 dB) response around the frequency range of interest (i.e., 0.9–1.1 GHz). Thus, it can be ensured that the feed will have a minimal effect on the radiation performance of the antenna and also will not introduce significant dispersion into the antenna response.

The loading inductance was achieved using a Coilcraft 0603HP-33N inductor, which has a nominal value of 35.45 nH at 1 GHz as determined from its manufacturer-provided SPICE model [39]. This value is deliberately chosen to be less than 42.97 nH used in the simulation of the miniaturized antenna so as to prove that this technique need not rely on the availability of a particular design value of lumped loading inductor, if the inductance may be otherwise compensated (as would be needed, for example, to compensate tolerances on off-the-shelf components and the effect of variations in landing patterns). Later in this section, it will be shown that any effect on the antenna $S_{11}$ response caused by the use of a smaller-valued inductance and the feed structure, indeed, can be neutralized simply by modifying the folded arm slightly and fine-tuning the loading capacitance.

The antenna was simulated in HFSS with the selected lumped inductor in the presence of the feed structure. It is important to note that every practical lumped inductor is associated with some amount of ohmic loss, which is usually quantified as the quality factor ($Q$) of the inductor and can be represented as an equivalent series resistance. The $Q$ of the selected inductor was found to be 79 at 1 GHz, suggesting an equivalent series resistance of 2.8 Ω, which was also introduced in the simulation. The corresponding $S_{11}$ response was compared with the return-loss response shown in Fig. 3(c). The comparison is shown in Fig. 6. As expected, the use of a smaller value of loading inductance shifts the peak return-loss value, in this case by 95 MHz. The return loss has also degraded by 9.64 dB, which occurs due to both the mismatch caused by a different loading reactance value and the added loss to the structure through the finite inductor $Q$.

However, as mentioned earlier, all these effects can be compensated and better matching can be realized at 1 GHz for the whole structure by parametrically tuning the different antenna parameters. For the selected lumped inductor, in order to bring the return-loss peak to 1 GHz and compensate for the lost inductance, two perpendicular sections of width $w_p = 1$ mm and length $l_p$ were added to the folded arm at the $L$-loading locations as shown in Fig. 7. The length of these additional sections was fine-tuned to provide the necessary additional inductive loading. In addition, the finger length ($C_l$) of the interdigitated capacitor was readjusted to vary the overall loading capacitance and re-establish the required proximity of $f_{AR,1}$ and $f_{R,N}$ around 1 GHz. All other antenna parameters were kept unchanged. It was found that, for $l_p = 2.11$ mm and $C_l = 0.64$ mm (approximate equivalent loading capacitance 102.9 fF), a much better return-loss response was obtained with $S_{11} = -31.78$ dB at 1 GHz and a 10 dB bandwidth of 15 MHz. The enhancement of bandwidth can be attributed largely to the finite losses introduced to the lumped loading inductors.

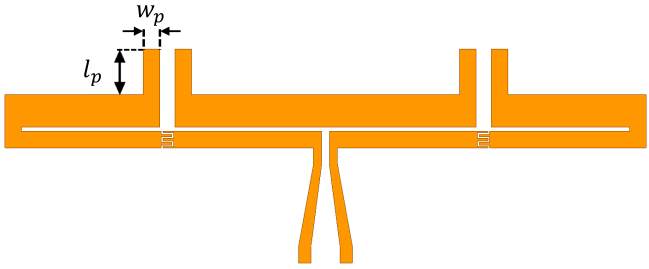

Fig. 7. Antenna layout with a modified folded arm.

The radiation efficiency is expected to be strongly reduced by the inductor $Q$ but is also susceptible to additional insertion losses incurred due to multiple scattering outside the antenna operating bandwidth, where the antenna is strongly mismatched to the feed. To quantify both sources of power dissipation, we may define an overall radiation efficiency of the antenna with the feed structure ($\eta_{rad,O}$), which may be estimated using the following equation:

$$\eta_{rad,O} = \eta_F \times \eta_{rad,A}. \quad (1)$$

Here, $\eta_F$ is the efficiency of the feed structure in transmitting the accepted power, and $\eta_{rad,A}$ is the radiation efficiency of the antenna including the parasitic effects of the feed layout, which can slightly shift the antenna resonance. This was obtained from HFSS by directly exciting the antenna in the presence of the feed layout, excluding the balun transformer. The feed efficiency $\eta_{rad,O}$ can be written in terms of the scattering parameters of the feed structure including the balun structure (subscript $F$) and the antenna (subscript $A$), as follows:

$$\eta_F = \frac{|S_{21,F}|^2 \times (1 - |S_{11,A}|^2)}{1 - \left|S_{11,F} + \frac{S_{12,F} S_{21,F} S_{11,A}}{1 - S_{22,F} S_{11,A}}\right|^2}. \quad (2)$$

The derivation of (2) is detailed in the Appendix.

Using (1), a comparison was made between the reactively loaded FD antenna and an unloaded FD antenna having the same overall footprint, both in the presence of the feed structure, and shown in Fig. 8. It can be easily seen that, although reduced due to the introduction of losses, the use of asymmetric loading to enable impedance matching nevertheless results in a substantially better radiation efficiency than the unloaded case. $\eta_{rad,O}$ of the loaded FD antenna was found to be 35.3% at 1 GHz, whereas the radiation efficiency is just 1.9% for the unloaded FD antenna at the same frequency. It is also worth comparing the simulated realized gains of the two antennas, which takes into account both the matching and the radiation efficiency, and which can be related more directly to the radiated power. It can be seen from Fig. 9 that, as expected, the loaded FD antenna has a much better realized broadside gain ($-3.80$ dB) than its unloaded counterpart ($-28.02$ dB) at 1 GHz.

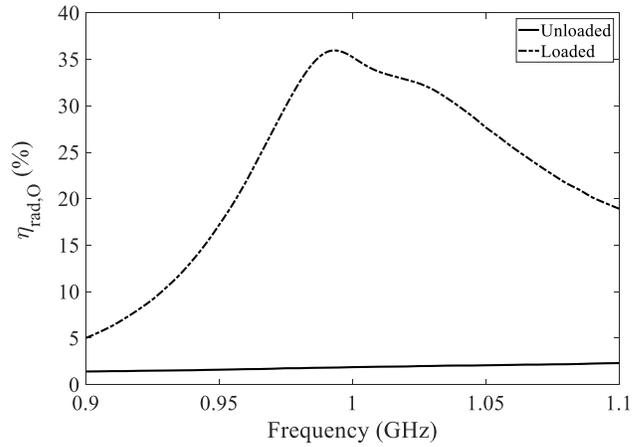

Fig. 8. Comparison of overall radiation efficiencies ($\eta_{rad,O}$) of unloaded and loaded miniaturized antennas, in the presence of the feed structure.

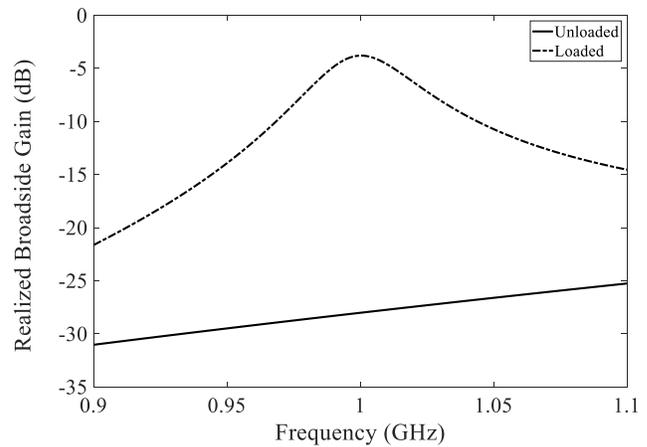

Fig. 9. Comparison of realized gains of unloaded and loaded miniaturized antennas in the broadside direction, in the presence of the feed structure.

Although not shown, the proposed miniaturized FD antenna was also compared with an equivalent (38 mm-long) miniaturized conventional dipole antenna, loaded inductively so as to make it resonant at $f_R = 1$ GHz. It was observed that, for equal inductor $Q$, both antennas demonstrated nearly identical radiation efficiency when fed using idealized matched sources. However, although the proposed miniaturized FD antenna could be practically designed for an essentially arbitrary input impedance suitable for matching to any balanced feed, the miniaturized conventional dipole would generally require an external matching network in order to produce comparable realized gain.

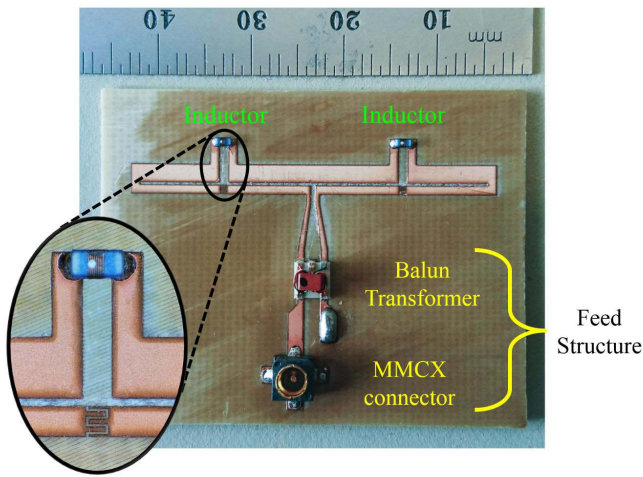

Fig. 10. Fabricated prototype of the loaded miniaturized FD antenna.

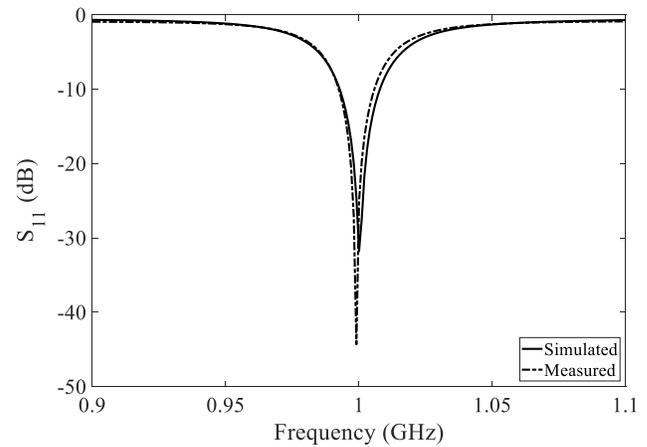

Fig. 11. Simulated versus measured $S_{11}$.

Finally, it is important to note that the observed $S_{11}$ bandwidth, radiation efficiency, and antenna quality factor of the miniaturized FD antenna are well within the Chu–Harrington limit as given in [40] and [41].

## III. FABRICATION AND MEASUREMENT

### A. Fabrication

The final parametrically tuned structure was fabricated using an LPKF ProtoLaser U3 milling machine that uses a laser to pattern designs on a dielectric substrate. The lumped inductors were soldered onto the antenna surface, along with the balun transformer and the MMCX surface-mount connector, and the vias were created using LPKF 0.9 mm copper rivets inserted into the substrate using a manual rivet punch and then soldered. The fabricated prototype is shown in Fig. 10.

### B. Measurement

The return-loss characteristics of the fabricated antenna were measured using a Keysight PNA-X (model N5244A) vector network analyzer. Early prototypes exhibited a matching frequency slightly higher than 1 GHz, suggesting that the inductance of the surface-mount inductor component used in this study was slightly less than its reported nominal value. Thus, the length $l_p$ of the perpendicular sections was further increased to 2.75 mm (keeping the other antenna parameters unchanged) in order to bring the return-loss peak closer to 1 GHz. Excellent −44.36 dB matching was obtained for the fabricated antenna at 0.999 GHz (∼75% miniaturized). It was found through simulations that a reduction in the loading inductance from its reported nominal value of 35.45 to 34.15 nH (which is within the specified ±5% tolerance value of the inductor) predicts the observed return-loss maximum at 0.999 GHz. The measured return loss was compared to the simulated result described in Section II-C ($l_p = 2.11$ mm, $L = 35.45$ nH) and shown in Fig. 11. Excellent agreement was found between the simulated and measured return-loss responses with a 10 dB return-loss bandwidth of approximately 15 MHz (1.5%).

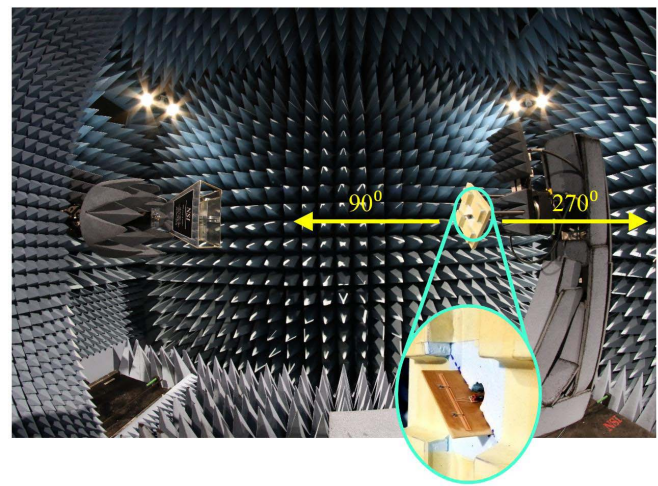

Fig. 12. Experimental setup for measuring the normalized radiation patterns (E-plane is the horizontal plane).

The radiation patterns were measured at 0.999 GHz inside a near-field anechoic chamber manufactured by NSI-MI Inc., using a dual-ridged horn antenna that has a bandwidth from 750 MHz to 10 GHz as the transmitting antenna, while the fabricated FD antenna was used as the receiver. Due to the lack of experimental capabilities to accurately measure gain, only the normalized radiation patterns have been presented. The measurement setup is shown in Fig. 12. The measured E- and H-plane patterns were compared with their simulated counterparts and a good agreement was found between them. The corresponding patterns are shown in Fig. 13. The measured co-pol to cross-pol separation at broadside direction (90°) is approximately 30 dB. It was observed from simulations that the added vertical segments at the L-loading locations have a minimal contribution to the cross-pol magnitudes. Once again, these data were obtained incorporating the parasitic effect of the feed layout on the antenna response, but without the balun transformer. As a result, the feed has only a very minimal effect on the simulated patterns. Thus, the observed difference between the simulated and measured patterns, particularly in the cross-pol magnitudes, can be attributed to the radiation from the actual currents on the feed structure of the fabricated antenna and also potentially multipath reflections from the measurement setup inside the chamber. Finally, the slight discrepancies evident in the E- and H-plane patterns at angles around 270° were caused due to the blockage created by the vertical mounting stand, as indicated in Fig. 12.

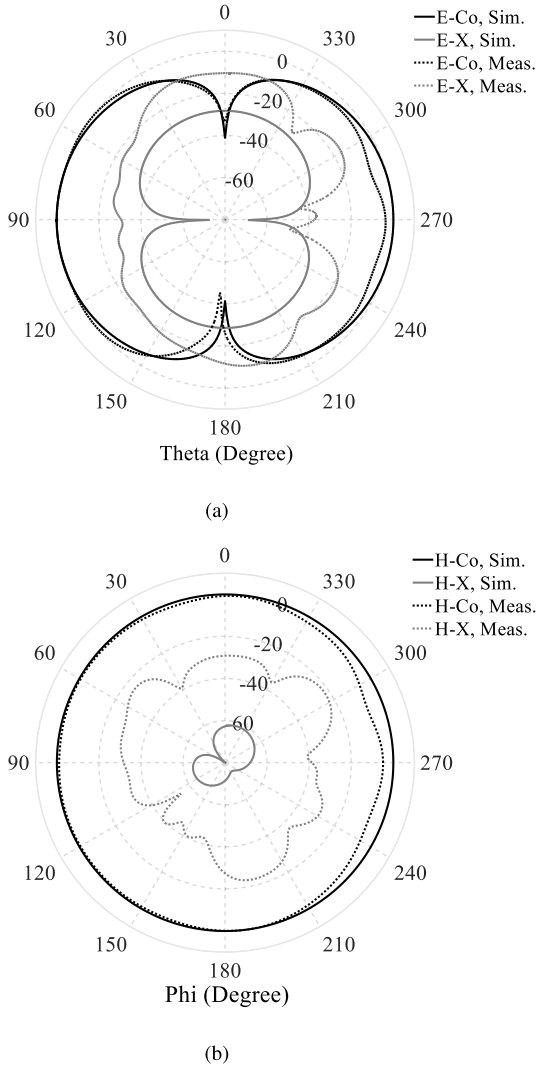

Fig. 13. Comparison of normalized radiation patterns at 0.999 GHz (simulated versus measured). (a) E-plane. (b) H-plane.

## IV. CONCLUSION

This article demonstrated that engineering the $Z_{in}$ of a printed FD antenna using asymmetric series-$LC$ loading and arm widths can enable excellent impedance matching, co- to cross-polarization separation, and improved radiation efficiency, even when it is highly miniaturized (∼75%) at the operating frequency. The miniaturized antenna does not require any external matching network and has attractive radiation performance suitable for sensor applications. Finally, the measured results are in excellent agreement with the simulated results.

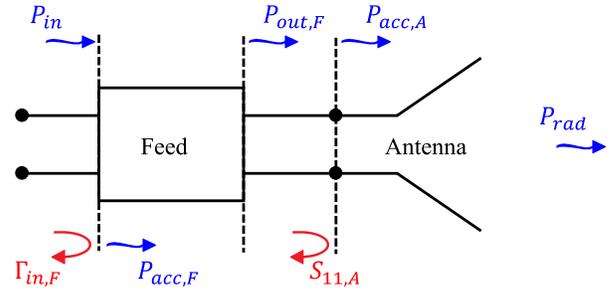

Fig. 14. Schematic of the antenna with the feed structure.

Although the approach presented in this article was used to achieve a real $Z_{in}$ at the target operating frequency, it may more generally be used to realize complex $Z_{in}$ for conjugate matching in some sensor applications like RFID, in which antennas are driven by chips possessing either an inductive or a capacitive output impedance. Furthermore, a fully printed equivalent antenna can be developed, which will not require any lumped components and, thus, will not be susceptible to the tolerances in off-the-shelf lumped component values, and will also be easier and less costly to fabricate. Additionally, the locations of reactive loading on the driven and folded arms can be varied in order to further improve the response.

## APPENDIX
## CALCULATION OF OVERALL RADIATION EFFICIENCY

The antenna with the feed structure can be represented using the schematic shown in Fig. 14, where $P_{in}$ is the overall input power, $P_{acc,F}$ is the power accepted into the feed structure when terminated in the antenna, $P_{out,F}$ is the power output from the feed structure, $P_{acc,A}$ is the power accepted into the antenna, and finally, $P_{rad}$ is the radiated power from the antenna.

Thus, the overall radiation efficiency of the antenna including the feed structure can be given by the following equation:

$$\eta_{rad,O} = \frac{P_{rad}}{P_{acc,F}} = \frac{P_{rad}}{P_{in}(1-|\Gamma_{in,F}|^2)}$$
$$= \frac{P_{rad}}{P_{in}\left(1 - \left|S_{11,F} + \frac{S_{12,F}S_{21,F}S_{11,A}}{1 - S_{22,F}S_{11,A}}\right|^2\right)} \quad (3)$$

where $S_{11,F}$, $S_{12,F}$, $S_{21,F}$, and $S_{22,F}$ are the generalized scattering parameters of the feed structure and $S_{11,A}$ pertains to the antenna loaded by the feed structure parasitically, but excluding the balun transformer.

Now, $P_{rad}$ can be given as follows:

$$P_{rad} = P_{acc,A} \times \eta_{rad,A} \quad (4)$$

where $\eta_{rad,A}$ is the radiation efficiency of the antenna including the parasitic effects of the feed layout, which was directly obtained from HFSS. The accepted power into the antenna $P_{acc,A}$ can be expressed using the following equation:

$$P_{acc,A} = P_{out,F} \times (1 - |S_{11,A}|^2)$$
$$= P_{in} \times |S_{21,F}|^2 \times (1 - |S_{11,A}|^2). \quad (5)$$

Thus, incorporating (4) and (5) into (3), the overall radiation efficiency can be obtained as

$$\eta_{rad,O} = \frac{P_{in} \times |S_{21,F}|^2 \times (1 - |S_{11,A}|^2)}{P_{in}\left(1 - \left|S_{11,F} + \frac{S_{12,F}S_{21,F}S_{11,A}}{1 - S_{22,F}S_{11,A}}\right|^2\right)} \times \eta_{rad,A}$$
$$= \eta_F \times \eta_{rad,A} \qquad (6)$$

where

$$\eta_F = \frac{|S_{21,F}|^2 \times (1 - |S_{11,A}|^2)}{1 - \left|S_{11,F} + \frac{S_{12,F}S_{21,F}S_{11,A}}{1 - S_{22,F}S_{11,A}}\right|^2}. \qquad (7)$$

It may also be noted that, for an inherently well-matched antenna, $S_{11,A} \approx 0$, hence (7) reduces to

$$\eta_F = \frac{|S_{21,F}|^2}{1 - |S_{11,F}|^2}. \qquad (8)$$

Furthermore, for a broadband, low loss, and well-matched feed structure, considering $S_{11,F} \approx S_{22,F} \approx 0$ and $S_{12,F} \approx S_{21,F} \approx 1$, $\eta_F$ and $\eta_{rad,O}$ can be obtained as

$$\eta_F = 1 \qquad (9)$$

and

$$\eta_{rad,O} = \eta_{rad,A}. \qquad (10)$$

## Acknowledgment

The authors would like to thank CMC Microsystems for providing simulation tools, Coilcraft Inc. for supplying surface-mount inductor samples, and the Canada Foundation for Innovation (CFI) and the Province of Alberta for experimental facilities.

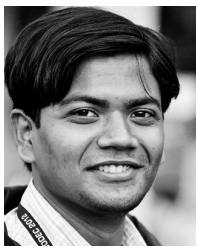

**Sanghamitro Das** (Member, IEEE) received the B.Sc. degree (Hons.) in physics and the B.Tech. and M.Tech. degrees in radio physics and electronics from the University of Calcutta, Kolkata, India, in 2008, 2011, and 2013, respectively, and the Ph.D. degree in electrical engineering from the University of Alberta, Edmonton, AB, Canada, in 2019.

He is currently working as a Research Assistant with the University of Alberta. His current research interests include miniaturized antennas for sensing and monitoring, novel miniaturization techniques for antennas as well as investigation of their time- and frequency-domain characteristics, and metamaterial flat lenses.

Dr. Das received the 3-year Alberta Innovates Graduate Student Scholarship in 2014, and the IEEE AP-S Doctoral Research Grant in 2016. He serves as an active volunteer for the IEEE Northern Canada Section AP-S/ MTT-S Joint Chapter.

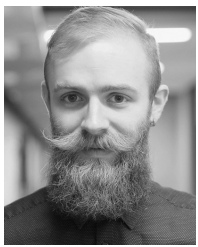

**David J. Sawyer** (Member, IEEE) received the B.Sc. degree in engineering physics and the M.Sc. degree in electrical engineering from the University of Alberta, Edmonton, AB, Canada, in 2016 and 2018, respectively.

He has worked on a variety of software projects with Edmonton-based company Sequiter Software from 2011 to 2015. He is currently an RF Specialist for the DVTEST Inc., Pickering, ON, Canada. His research interests include fundamental limits in antenna design, imaging technologies, and computational methods in electromagnetics.

Mr. Sawyer received the Faculty of Engineering and University of Alberta Academic Excellence Scholarships, the Deans Research Award, and the Queen Elizabeth II Scholarship.

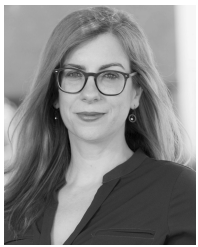

**Nectaria Diamanti** received the B.Sc. degree (Hons.) in geology and the M.Sc. degree in geophysics from the Aristotle University of Thessaloniki, Thessaloniki, Greece, in 2000 and 2002, respectively, and the Ph.D. degree in engineering and electronics from the University of Edinburgh, Edinburgh, U.K., in 2008.

She is currently a Research Scientist with the Department of Geophysics, Aristotle University of Thessaloniki. From 2013 to 2018, she worked for Sensors & Software Inc., Mississauga, ON, Canada, in ground-penetrating radar (GPR) research and development, and she continues to be an active contributor to their scientific research and applications advancements as well as their outreach and cooperative projects. Her main research activity involves the development of geophysical techniques, especially GPR technologies and their application to geophysical/engineering problems ranging from environmental monitoring to nondestructive testing and archeological prospection. Her areas of research include numerical modeling using the finite-difference time-domain (FDTD) technique and application of numerical modeling to GPR.

Dr. Diamanti is a member of the Society of Exploration Geophysicists (SEG) and European Association of Geoscientists and Engineers (EAGE). Since 2016, she has been an Associate Editor for GPR in the *Journal of Environmental & Engineering Geophysics* (JEEG).

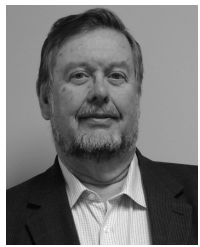

**A. Peter Annan** received the B.A.Sc. degree in engineering science and the M.Sc. degree in geophysics from the University of Toronto, Toronto, ON, Canada, in 1968 and 1970, respectively, and the Ph.D. degree in physics from the Memorial University of Newfoundland, St. John's, ON, in 1974.

He is currently the CEO of Sensors & Software Inc., Mississauga, ON, a GPR Instrument Company. He has a long history of pioneering GPR advancement including survey applications, instrument development and product commercialization. He has extensive experience in geophysical survey design, operation and project management plus an active interest in airborne electromagnetic geophysical system design and commercialization.

Dr. Annan is a member of the SEG, ASEG, EAEG, EEGS, PEO, CGU, and a number of other professional associations.

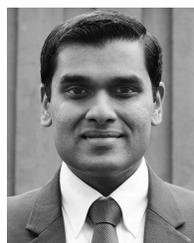

**Ashwin K. Iyer** (Senior Member, IEEE) received the B.A.Sc., M.A.Sc., and Ph.D. degrees in electrical engineering from the University of Toronto, Toronto, ON, Canada, in 2001, 2003, and 2009, respectively, with a focus on the discovery and development of the negative-refractive-index transmission-line approach to metamaterial design and the realization of metamaterial lenses for free-space microwave subdiffraction imaging.

He is currently an Associate Professor with the Department of Electrical and Computer Engineering, University of Alberta, Edmonton, AB, Canada, where he leads a team of graduate students investigating novel RF/microwave circuits and techniques, fundamental electromagnetic theory, antennas, and engineered metamaterials, with an emphasis on their applications to microwave and optical devices, defense technologies, and biomedicine. He has coauthored a number of highly cited articles and book chapters on the subject of metamaterials.

Dr. Iyer is a member of the IEEE AP-S Education Committee and a Registered Member of the Association of Professional Engineers and Geoscientists of Alberta. He was a recipient of the IEEE AP-S R. W. P. King Award in 2008, the IEEE AP-S Donald G. Dudley Jr. Undergraduate Teaching Award in 2015, the University of Alberta Provost's Award for Early Achievement of Excellence in Undergraduate Teaching in 2014, and the University of Alberta Rutherford Award for Excellence in Undergraduate Teaching in 2018. His students are the recipients of several major national and international awards for their research. He serves as the Co-Chair for the IEEE Northern Canada Section's award-winning Joint Chapter of the AP-S and MTT-S Societies and a Technical Program Committee Co-Chair for the 2020 AP-S/URSI International Symposium. From 2012 to 2018, he was an Associate Editor of the IEEE TRANSACTIONS ON ANTENNAS AND PROPAGATION and currently serves as a Track Editor. He is the Guest Editor of the IEEE TRANSACTIONS ON ANTENNAS AND PROPAGATION special issue on Recent Advances in Metamaterials and Metasurfaces.